\documentclass [a4paper,fleqn]{article}

\usepackage {amsmath} \usepackage{amssymb} \usepackage{cite} \usepackage{amsthm}
\numberwithin{equation}{section} \numberwithin{table}{section} \mathindent=0pt
\theoremstyle{plain} 

\numberwithin{theorem}{section}

\begin{document}

\title{Solitary and periodic solutions of the generalized Kuramoto - Sivashinsky equation}
\author{Nikolai A. Kudryashov}
\date{Department of Applied Mathematics\\
Moscow  Engineering and Physics Institute\\
(State university)\\
31 Kashirskoe Shosse,  115409\\
Moscow, Russian Federation} \maketitle

\begin{abstract}

The generalized Kuramoto - Sivashinsky in the case of the power
nonlinearity with arbitrary degree is considered. New exact
solutions of this equation are presented.

\end{abstract}



\section{Introduction}

In this paper we present exact solutions of the generalized Kuramoto
- Sivashinsky equation
\begin{equation}\begin{gathered}
\label{0.1}u_t+u^m\,u_x+\alpha\,u_{xx}+\beta\,u_{xxx}+\gamma\,u_{xxxx}=0
\end{gathered}
\end{equation}
Nonlinear evolution equation \eqref{0.1} at $m=1$ has been studied
by a number of authors from various viewpoints. This equation has
drown much attention not only because it is interesting as a simple
one - dimensional nonlinear evolution equation including effects of
instability, dissipation and dispersion but also it is important for
description in engineering and scientific problems. Equation
\eqref{0.1} in work \cite{Kuramoto01} was used for explanation of
the origin of persistent wave propagation through medium of reaction
- diffusion type. In paper \cite{Sivashinsky01} equation \eqref{0.1}
was obtained at $m=1$ for description of the nonlinear evolution of
the disturbed flame front. We can meet the application of equation
\eqref{0.1} for studying of motion of a viscous incompressible
flowing down an inclined plane \cite{Benney01, Topper01, Shkadov01}.
Mathematical model for consideration dissipative waves in plasma
physics  by means of equation \eqref{0.1} was presented in
\cite{Cohen01}. Elementary particles as solutions of the Kuramoto -
Sivashinsky equation were studied in \cite{Michelson03}. Equation
\eqref{0.1} at $m\neq1$ also can be used for the description of
physical applications. For a example this equation were derived for
description of nonlinear long waves in viscous - elastic tube
\cite{Kudryashov08}.

Exact solution of equation \eqref{1.2} in the form of the solitary
wave at $m=1$ and at $\beta=0$ were obtained in \cite{Kuramoto01,
Conte01}.  Solitary wave solutions of equation \eqref{1.2} at $m=1$
in the case $\beta\neq0$ were found in works \cite{Kudryashov88,
Kudryashov91, Kudryashov96} for cases:
\begin{equation}\begin{gathered}
\label{0.2}\alpha=\frac{\beta^2}{16\,\gamma}\quad
\alpha=\frac{47\,\beta^2}{144\,\gamma},\quad
\alpha=\frac{73\,\beta^2}{256\,\gamma}.
\end{gathered}
\end{equation}

Other forms of these solutions  were presented in works
\cite{Berloff01, Fu01, Zhang01, Khuri01, Nickel01}. Periodical
solutions of equation \eqref{1.2} were found in \cite{Kudryashov90,
Kudryashov90s} at $\alpha=\frac{\beta^2}{16\,\gamma}$.

Exact solutions of equation \eqref{1.2} at $m=2$ recently were
obtained \cite{Kudryashov08} for the four cases:
\begin{equation}\begin{gathered}
\label{0.3} \alpha=\frac{2\,\beta^2}{25\,\gamma}\quad
\alpha=\frac{71\,\beta^2}{225\,\gamma},\quad
\alpha=\frac{121\,\beta^2}{55\,\gamma},\quad
\alpha=\frac{374\,\beta^2}{2025\,\gamma}.
\end{gathered}\end{equation}

The aim of this paper is to search for exact solutions of equation
\eqref{0.1} in general case of arbitrary degree $m$. We present
exact solutions of equation \eqref{0.1} at  $m\neq -1$, $m\neq -3$
and $m\neq 0$. Some of these solutions are new.

\section{Solitary waves of equation \eqref{0.1}}

Using traveling wave
\begin{equation}\begin{gathered}
\label{1.1a}u(x,t)=w(z),\,\qquad z=x-C_0\,t
\end{gathered}
\end{equation}
from equation \eqref{0.1} we have
\begin{equation}\begin{gathered}
\label{1.2}w_{zzz}+\sigma\,w_{zz}+w_{z}-C_0\,w+\frac{1}{m+1}\,w^{m+1}=0,
\qquad m\neq -1
\end{gathered}
\end{equation}

Assuming in \eqref{1.2}
\begin{equation}\begin{gathered}
\label{1.3}w(z)=v(z)^{\frac{1}{m}}
\end{gathered}
\end{equation}
we have equation in the form

\begin{equation}\begin{gathered}
\label{1.4}\gamma\,{m}^{3}\,{v}^{2}\,v_{{{zzz}}}+\gamma\,{m}^{2}\,{v}^{2}\,v_{{{zzz}}}+
3\,\gamma\,m\,v\,v_{{z}}\,v_{{{zz}}}+\beta\,{m}^{3}\,{v}^{2}\,v_{{{zz}}}-
2\,\gamma\,m\,{v_{{z}}}^{3}-\\
\\
-3\,\gamma\,{m}^{3}v\,v_{{z}}\,v_{{{zz}}}
+\gamma\,{v_{{z}}}^{3}-\gamma\,{m}^{2}\,{v_{{z}}}^{3}+2\,\gamma\,{m}^{3}\,{v_{{z}}}^{3}+
\beta\,m\,v\,{v_{{z}}}^{2}
-\\
\\
-\beta\,{m}^{3}\,v\,{v_{{z}}}^{2}+\beta\,{m}^{2}\,{v}^{2}\,v_{{{zz}}}+
\alpha\,{m}^{3}\,{v}^{2}\,v_{{z}}+\alpha\,{m}^{2}\,{v}^{2}\,v_{{z}}+{m}^{3}\,{v}
^{4}-\\
\\
-{C_0}\,{m}^{4}\,{v}^{3}-{C_0}\,{m}^{3}\,{v}^{3}=0
\end{gathered}
\end{equation}

Study of analytical properties of equation \eqref{1.4} allows us to
determine that in the general case the meromorphic solutions of
equation \eqref{1.4} can be found  taking into account two cases:
\begin{equation}\begin{gathered}
\label{1.4b} \alpha={\frac {{\beta}^{2}\,m\, }{\gamma \left( m+3
\right) ^{2}}},\qquad \alpha=\,{\frac {{\beta}^{2}\,\left(
2\,{m}^{2}+18\,m+27 \right)\, }{9\,\gamma
 \left( m+3 \right) ^{2}}}
\end{gathered}\end{equation}
Consider the first case. The pole order of the solution of equation
\eqref{1.4} is equal to three therefore we look for exact solutions
of equation \eqref{1.4} in the form
\begin{equation}\begin{gathered}
\label{1.5}v(z)=A_0+A_1\,Y(z)+A_2\,Y(z)^2+A_3\,Y(z)^3,
\end{gathered}
\end{equation}
where $Y(z)$ is solution of equation with the first order pole. Let
us take simplest equation in the form \cite{Kudryashov05a,
Kudryashov05b}
\begin{equation}\begin{gathered}
\label{1.5a}Y_z=-Y^2+b.
\end{gathered}
\end{equation}
Substituting \eqref{1.5} into \eqref{1.4} and taking into account
equation \eqref{1.5a} we have
\begin{equation}\begin{gathered}
\label{1.6}b\,=\,{\frac {{\beta}^{2}{m}^{2}}{4\,{\gamma}^{2} \left(
m+3 \right) ^{2}}}.
\end{gathered}
\end{equation}

We have also the following values of coefficients in \eqref{1.5}
\begin{equation}\begin{gathered}
\label{1.7}A_{{3}}={\frac {3\,\gamma \left( 2\,m+3 \right)  \left(
m+3 \right) \left( m+1 \right) }{{m}^{3}}},\\
\\
A_{{2}}=-\,{\frac {3\, \left( 2\,m+3 \right)  \left( m+1 \right)
\beta }{2\,{m}^{2}}},\quad A_{{1}}=-\,{\frac {3\, \left( 2\,m+3
\right)  \left( m+1 \right) { \beta}^{2}}{4\, \left( m+3 \right)
m\,\gamma}},\\
\\
C_{{0}}=\,{\frac {2\, \left( m+2 \right) {\beta}^{3}}{{\gamma}^{2}
\left( m+3
 \right) ^{3}}}, \quad A_0=\,{\frac {3\, \left( 2\,m+3 \right)  \left( m+1
\right) {\beta}^{3}}{{8\, \gamma}^{2} \left( m+3 \right) ^{2}}}.
\end{gathered}
\end{equation}

We have solutions of equation \eqref{1.5a} in the form
\begin{equation}\begin{gathered}
\label{1.12}Y(z)\,=\,\frac{\beta\,m}{2\,{\gamma}\,\left( m+3
\right)}\,\,\tanh \left\{ \,{\frac { \beta\,m\,\left( z-{z_0}
\right) }{2\,\gamma\, \left( m+3 \right) }} \right\}.
\end{gathered}
\end{equation}
Solution of equation \eqref{1.2} takes the form
\begin{equation}\begin{gathered}
\label{1.13}w(z)=\left(\frac{3\left( 2\,m+3 \right) \left( m+1
\right) { \beta}^{3}}{8\,\gamma\,\left( m+3
\right)}\right)^{\frac1m}\,\,\left( 1-\tanh \left\{ {\frac {
\beta\,m\, \left( z-{z_0}
 \right)}{2\,\gamma \left( m+3 \right) }} \right\} \right) ^{\frac2m}
\\
 \left( 1+\tanh \left\{ {\frac { \beta\, m\,\left( z-{z_0} \right)
}{2\,\gamma \left( m+3 \right) }}\right\}\right)^{\frac1m},\quad
m\neq 0, \quad m\neq -3,\quad m\neq -1.
\end{gathered}
\end{equation}

This is new solution of equation \eqref{0.1}. Solution \eqref{1.13}
at $m=1$ takes the form of the solitary wave that was obtained in
\cite{Kudryashov88}. In the case $m=2$ this solution corresponds to
exact solution of equation \eqref{0.1} published in \cite
{Kudryashov08}.

Consider the second case. Solutions of equation \eqref{1.4} we look
for taking into consideration the transformation \eqref{1.5} again.
Function $Y(z)$ satisfies equation \eqref{1.6} as well. In this case
we obtain the expression for the parameter $b$ in the form
\begin{equation}\begin{gathered}
\label{1.14}b\,=\,{\frac {{\beta}^{2}{m}^{2}}{
36\,{\gamma}^{2}\,\left( m+3 \right) ^{2}}}.
\end{gathered}
\end{equation}
We also obtain coefficients $A_3$, $A_2$, $A_1$, $C_0$ and $A_0$ as
following
\begin{equation}\begin{gathered}
\label{1.15}A_{{3}}=\,{\frac {3\,\gamma\, \left( 2\,m+3 \right)
\left( m+3 \right) \left( m+1 \right) }{{m}^{3}}},\\
\\
A_{{2}}=-\,{\frac {3\, \left( 2\,m+3 \right)  \left( m+1 \right)
\beta }{2\,\,{m}^{2}}}, \quad A_{{1}}=\,{\frac { \left( 2\,m+3
\right)  \left( m+1 \right) {\beta }^{2}}{4\,\gamma\,m\, \left( m+3
\right)}},\\
\\
C_{{0}}=-\,{\frac { \left( 2\,m+3 \right) {\beta}^{3}}{ 9\,\left(
m+3 \right) ^{2}{\gamma}^{2}}},\quad A_{{0}}=-\,{\frac {{\beta}^{3}
\left( 2\,m+3 \right) \left( m+1 \right) }{72\, \left( m+3 \right)
^{2}{\gamma}^{2}}}.
\end{gathered}
\end{equation}

Using these coefficients we have solution of equation \eqref{1.2}.
It takes the form
\begin{equation}\begin{gathered}
\label{1.19}w(z)=\left({\frac {{\beta}^{3}\,\left( 2\,m+3
\right)\,\left( m+1 \right) }{72\,\gamma^2\,\left( m+3
\right)^2\,}}\right)^{\frac1m}\,
 \left( \tanh \left\{ {\frac {\beta\,m\,
 \left( z-{z_0} \right) }{6\,\gamma\, \left( m+3 \right) }} \right\} -
1 \right) ^{\frac3m},\\
\\
m\neq 0,\quad m\neq-1,\quad m\neq-3.
\end{gathered}
\end{equation}
Other forms of this solution were found in works \cite{Zhu01,
Kudryashov07, Qin01}. At $m=1$ from \eqref{1.19} we have one of the
solitary solutions of work \cite{Kudryashov88}. In the case $m=2$ we
get solution \cite{Kudryashov08}.

\section{Periodical waves of equation \eqref{0.1}}

Let us find periodical solutions of equation \eqref{1.2} taking into
consideration the formula \cite{Kudryashov90}
\begin{equation}\begin{gathered}
\label{2.1}w(z)=A_{0}+A_{1}\,R+A_{2}\,R_{z},\qquad R\equiv R(z),
\end{gathered}
\end{equation}
taking into account the relation for the parameter $\alpha$
\begin{equation}\begin{gathered}
\label{2.1a}\alpha\,=\,{\frac {{\beta}^{2}\,m}{\gamma \left( m+3
\right) ^{2}}}.
\end{gathered}
\end{equation}

We assume that $R(z)$ is solution of equation for the Weierstrass
function
\begin{equation}\begin{gathered}
\label{2.2}R_{{{zz}}}+6\,{R}^{2}-a\,R-b=0
\end{gathered}
\end{equation}
From equation \eqref{2.2} we have equation
\begin{equation}\begin{gathered}
\label{2.2a}R_{{{z}}}^2+4\,{R}^{3}-a\,R^2-2\,b\,R+d=0
\end{gathered}
\end{equation}

Substituting \eqref{2.1} into equation \eqref{1.3} and taking into
account equations \eqref{2.2} and \eqref{2.2a} we have
\begin{equation}\begin{gathered}
\label{2.4}b=\,{\frac {{m}^{4}{\beta}^{4}-
\gamma^4\,a^2\,\left(m+3\right)^2\,}{{24\,\gamma}^{4} \left( m+3
\right) ^{4}}},\quad A_2==\,{\frac {3\,\gamma \left( 2\,m+3 \right)
\left(
m+3 \right) \left( m+1 \right) }{2\,{m}^{3}}},\\
\\
A_{{1}}=\,{\frac {3 \,\beta\, \left( 2\,m+3 \right) \left( m+1
\right) } {2\,{m}^{2}}}, \quad C_0=\,{\frac { 2\,{\beta}^{3}\,\left(
m+2 \right) }{{\gamma}^{2} \left( m+3 \right) ^{3}}},\\
\\
A_0=\,{\frac { \beta\,\left( 2\,m+3 \right) \, \left( m+1
\right)\,\left({\beta}^{2}\,{m}^{2}-{\gamma}^{2}\,({m+3})^{2}\,a\,\right)
}{{8\,\gamma}^{2}{m}^{2} \left( m+3 \right) ^{2}}},
\end{gathered}
\end{equation}
\begin{equation}\begin{gathered}
\label{2.8}d=\,{\frac {\left(
2\,{\beta}^{2}{m}^{2}+{\gamma}^{2}\,(m+3)^{2}\,a\,\right)\, \left(
{\beta}^{2}{m}^{2}-{\gamma}^{2}\, {(m+3)}^{2}\,a\, \right)
^{2}}{432\,\,{\gamma}^{6} \left( m+3 \right) ^{ 6}}}
\end{gathered}
\end{equation}

Solution of equation \eqref{1.2} in this case takes the form
\begin{equation}\begin{gathered}
\label{2.9} w(z)=\left(\frac{3\,\left( m+1 \right)\,\left( 2\,m+3
\right)}{2\,{m}^{2}} \,\right)^{\frac1m}
\\
 \left( \beta\,R \left( z \right) +{\frac {\gamma
\left( m+3 \right) }{m}}\,R_z +\,{\frac {
\beta\,\left({\beta}^{2}{m}^{2}-{\gamma}^{2}\, \left( m+3 \right)
^{2} a\,\right)}{12\,{\gamma}^{2} \left( m+3 \right) ^{2}}}
\right)^{\frac1m},
\end{gathered}\end{equation}
where $R(z)$ satisfies equation
\begin{equation}\begin{gathered}
\label{2.10}{R_{{z}}}^{2}+4\,{R}^{3}-a\,{R}^{2}-\,{\frac { \left(
{m}^{4}{\beta} ^{4}-{\gamma}^{4}\,{a}^{2} \,\left( m+3 \right) ^{4}
\right) }{12\,{\gamma}^{4} \left( m+3 \right) ^{4}}}\,\,R -\\
\\
-\,{\frac {\left(
2\,{\beta}^{2}{m}^{2}+{\gamma}^{2}\,(m+3)^{2}\,a\,\right)\, \left(
{\beta}^{2}{m}^{2}-{\gamma}^{2}\, {(m+3)}^{2}\,a\, \right)
^{2}}{432\,\,{\gamma}^{6} \left( m+3 \right) ^{ 6}}}=0
\end{gathered}
\end{equation}
In general case solutions of equation \eqref{2.10} expressed via
hyperbolic function because the algebraic equation
\begin{equation}\begin{gathered}
\label{2.11}4\,{R}^{3}-a\,{R}^{2}-\,{\frac { \left( {m}^{4}{\beta}
^{4}-{\gamma}^{4}\,{a}^{2} \,\left( m+3 \right) ^{4} \right)
}{12\,{\gamma}^{4} \left( m+3 \right) ^{4}}}\,\,R -\\
\\
\,{\frac {\left(
2\,{\beta}^{2}{m}^{2}+{\gamma}^{2}\,(m+3)^{2}\,a\,\right)\, \left(
{\beta}^{2}{m}^{2}-{\gamma}^{2}\, {(m+3)}^{2}\,a\, \right)
^{2}}{432\,\,{\gamma}^{6} \left( m+3 \right) ^{ 6}}}=0
\end{gathered}
\end{equation}
has two equal roots.

However there are periodic solutions of equation \eqref{0.1} at
$m=1$ and $m=3$. Parameter $C_0$ in these cases will be arbitrary
constant. Other coefficients can be found from expressions
\eqref{2.4}. Let us present these solutions using formula
\eqref{2.1}.

In the case $m=1$ we have $\alpha=\frac{\beta^2}{16\,\gamma}$.
Solution of equation \eqref{0.1} takes the form \cite{Kudryashov90}
\begin{equation}\begin{gathered}
\label{2.12}y(z)={ C_0}-\frac54\,\beta\, a-{\frac {1}{64}}\,{\frac
{{\beta}^{3}}{{\gamma}^{2}}}+15\,\beta\,\,R+60\,\gamma\,\,{R_z}
\end{gathered}
\end{equation}
where $R(z)$ satisfies equation for the Weierstrass function
\begin{equation}\begin{gathered}
\label{2.13}{R_{{z}}}^{2}+4\,{R}^{3}-a{R}^{2}-2\, \left(\,{\frac
{{\beta}^{4}}{6144\,{\gamma}^{4}}}-\frac{{a}^{2}}{24}\,
\right) R+\\
\\
+{\frac {13\,{\beta}^{6}}{4423680\,\gamma^{6}}}\,+\,\,{\frac
{a\,{\beta}^ {4}}{{36864\,\gamma}^{4}}}-\,{\frac
{{C_{{0}}}^{2}}{2160\,{\gamma}^{2}}}-{ \frac {{a}^{3}}{432}}\,=0
\end{gathered}
\end{equation}

In the case $m=3$ we have $\alpha=\frac{\beta^2}{12\,\gamma}$.
Periodic solution of equation \eqref{0.1} in this case can be
written in the form
\begin{equation}\begin{gathered}
\label{2.14}y(z)=\left(3\,{C_0}-\frac12\,\beta\,a-{\frac
{1}{72}}\,{\frac {{\beta}^{3}}{{\gamma}^{2
}}}+6\,\beta\,R+12\,\gamma\,R_{{z}} \right)^{\frac13}
\end{gathered}
\end{equation}
where $R(z)$ satisfies equation for the Weierstrass function in the
form
\begin{equation}\begin{gathered}
\label{2.15}{R_{{z}}}^{2}+4\,{R}^{3}-a\,{R}^{2}-2\, \left(\,{\frac
{C_{{0}}\beta}{8\,{\gamma}^{2}}} -{\frac
{11\,{\beta}^{4}}{3456\,{\gamma}^{4}}}-\frac{{a}^{2}}{24}\, \right) R-\\
\\
-{\frac {11\,{\beta}^{4}a}{20736\,{\gamma}^{ 4}}}+{\frac
{13\,{\beta}^{6}}{373248\,{\gamma}^{6}}}+{\frac
{C_{{0}}{\beta}^{3}}{{1728\,\gamma}^{4}}}+{\frac {C_{{0}}
\beta\,a}{48\,{\gamma}^{2}}}-{\frac
{{C_{{0}}}^{2}}{16\,{\gamma}^{2}}}-{\frac {{a}^{3}}{ 432}}\,=0
\end{gathered}
\end{equation}
This solution of equation \eqref{0.1} is a new. Let us note that
periodical solutions of equation \eqref{0.1} are appeared at $m=1$
and at $m=3$ when there is meromorphic solutions in the form of
solitary waves of equation \eqref{0.1}.

\end{document}